\documentclass[conference]{IEEEtran}
\IEEEoverridecommandlockouts

\usepackage{amsmath}
\usepackage{amsfonts}
\usepackage{amssymb}
\usepackage{graphicx}
\usepackage{epsfig}
\usepackage{subfig}
\usepackage{float}
\usepackage{epstopdf}
\usepackage{array}
\usepackage{multirow}
\usepackage{setspace}
\usepackage{color}
\usepackage[section]{placeins}
\usepackage{url}
\usepackage{algorithm}
\usepackage{algpseudocode}
\usepackage{multicol}
\usepackage{lipsum}
\usepackage{soul}
\usepackage{comment}
\usepackage{balance}
\usepackage{xcolor}
\usepackage{cite}
\usepackage[normalem]{ulem}
\usepackage[acronym]{glossaries}

\newtheorem{remark}{\textbf{Remark}}

\title{New Radio Physical Layer Abstraction for System-Level Simulations of 5G Networks}

\author{
  \IEEEauthorblockN{
    Sandra Lagen\IEEEauthorrefmark{1},~Kevin Wanuga\IEEEauthorrefmark{2},~Hussain Elkotby\IEEEauthorrefmark{2},~Sanjay Goyal\IEEEauthorrefmark{2},~Natale Patriciello\IEEEauthorrefmark{1},~Lorenza Giupponi\IEEEauthorrefmark{1}
   }
    
    \IEEEauthorblockA{\IEEEauthorrefmark{1}Centre Tecnol\`ogic de Telecomunicacions de Catalunya (CTTC/CERCA), Castelldefels, Barcelona, Spain\\ }

	\IEEEauthorblockA{\IEEEauthorrefmark{2}InterDigital Communications, Inc., Melville, New York, USA \\
	emails: \{slagen, npatriciello, lgiupponi\}@cttc.es, \{kevin.wanuga, hussain.elkotby, sanjay.goyal\}@interdigital.com 
	}
}

\begin{document}
\maketitle

\begin{abstract}
A physical layer (PHY) abstraction model estimates the PHY performance in system-level simulators to speed up the simulations. This paper presents a PHY abstraction model for 5G New Radio (NR) and its integration into an open-source ns-3 based NR system-level simulator. The model capitalizes on the exponential effective signal-to-interference-plus-noise ratio (SINR) mapping (EESM) and considers the latest NR specification. To generate it, we used an NR-compliant link-level simulator to calibrate the EESM method as well as to obtain SINR-block error rate (BLER) lookup tables for various NR configurations. We also illustrate the usability of the developed model through end-to-end simulations in ns-3, under different NR settings of modulation and coding schemes, hybrid automatic repeat request combining methods, and link adaptation approaches.
\end{abstract}


\IEEEpeerreviewmaketitle

\section{Introduction}
\label{sec:intro}
The 5G NR access technology includes
flexible, scalable, and forward-compatible PHY and medium access control (MAC) layers to support a broad range of center carrier frequencies, deployment options, and use cases~\cite{TS38300}. 
As compared to 4G Long Term Evolution (LTE), NR includes multiple new features such as flexible frame structures by means of various numerologies, support for wide channel bandwidth operations at millimeter-wave frequency bands, symbol-level scheduling through mini-slots of variable length, and new channel coding schemes (i.e., Low-Density Parity Check (LDPC) for data channels and Polar Codes for control channels). 
All the features and procedures of NR have been standardized by the 3rd Generation Partnership Project (3GPP), which has released the first phase of NR specification (as Release 15) in mid 2019 and the second phase (as Release 16) in mid 2020. 
Based on typical timings, the NR based 5G deployment and commercialization of 
Release 16 features is expected to start around 2022. Before entering into real implementations, it is fundamental to validate the performance of NR based systems.

Usually, researchers divide the wireless network simulators into two categories: link-level simulators (LLSs) and system-level simulators (SLSs). In the former, the simulations are performed by using a detailed and computationally intensive model of PHY processing of the air interface, such as channel coding-decoding, multi-antenna gains, and orthogonal frequency division multiplexing (OFDM) modulation. In the latter, the focus is on network-based analysis, including for example resource allocation, mobility, and interference management. In SLS, since simulations typically involve many nodes, the modeling of PHY processing is replaced by an abstraction model, implemented through an interface that is known as link-to-system mapping (L2SM)~\cite{mezzavilla:12, patidar:17}. The L2SM is a process that estimates the BLER of a MAC transport block (denoted as \textit{transport BLER} for short) as a function of the radio channel from pre-computed tables, without including execution of PHY processing, 
hence saving execution time.

In particular, for system-level simulations of multi-carrier based OFDM systems in frequency selective channels, the L2SM includes two key blocks: 1) the compression of the given set of post-processing SINRs experienced by the receiver over every sub-carrier, as reported from the channel model, into a single scalar value (called \textit{effective SINR}) and 2)
the computation of the transport BLER corresponding to the derived effective SINR, by using an appropriate SINR-BLER lookup table. 
The technique for SINR compression is known as effective SINR mapping (ESM). ESM uses a mapping function to obtain an effective SINR that summarizes the effect of multiple SINRs affecting the packet by modeling the link as an equivalent additive white Gaussian noise (AWGN) channel. 
Specifically, for each modulation and coding scheme (MCS), the mapping function parameters need to be calibrated such that the variation of SINRs is mapped to an effective SINR value that would produce the same BLER performance under an AWGN channel, as the experimental BLER that is measured in a fading channel with the same MCS~\cite{cipriano:08}. 

ESM has been widely used in SLSs of 4G LTE~\cite{hanzaz:11,mezzavilla:12}, IEEE 802.16 WiMAX~\cite{hanzaz:11b}, IEEE 802.11 Wi-Fi OFDM~\cite{patidar:17}, and recently 5G NR~\cite{s19051196}.
There are multiple models available for ESM~\cite{brueninghaus:05}, depending on the mapping function that is used: exponential ESM (EESM), mutual information ESM (MIESM), capacity ESM, and logarithmic ESM. There is no clear consensus among researchers regarding the selection of a single method. 
In IEEE 802.11, a variant of the MIESM method, known as received bit mutual information rate (RBIR)~\cite{wan:06}, is agreed to be used for SLSs~\cite{80211-14}. On the other hand, 3GPP recommends an EESM method characterized by a single parameter~\cite{TR25892}, for which the details are given in Section~\ref{sec:L2SM}.

The PHY abstraction of NR based systems is a complex task due to the multiple new features added to NR. In NR, in addition to number of resource blocks (RBs), the number of OFDM symbols can also be variably allocated to a user, which in combination with wide-bandwidth operation significantly increases the number of supported transport block sizes (TBSs). The inclusion of LDPC coding with multiple lifting sizes and two types of base graphs increases the complexity of the code block segmentation procedure at PHY. Moreover, NR supports various configurations for MCS tables, and modulation orders up to 256-QAM. All these features need to be considered to model NR performance appropriately.

In this paper, we present a PHY abstraction model for simulation of NR based 5G networks. The proposed model uses the EESM method, for which the optimal effective SINR mapping parameters are derived with a calibration procedure using an NR-compliant LLS. The NR-compliant LLS is also used to generate the SINR-BLER curves for various settings (e.g., different MCSs, MCS tables, and resource allocations), which are used to find the mapped transport BLER in SLSs. Thanks to the calibration procedure, we show that the L2SM for NR is insensitive to the NR numerology, which has not been found out before in the literature. Finally, the proposed model is integrated and validated through the NR SLS~\cite{5GLENA} of an open-source network simulator, i.e., ns-3. 
The code is publicly available to the research community (https://5g-lena.cttc.es), as a new interface of the ns-3 NR SLS~\cite{5GLENA}. Therefore, it can be tested and evaluated  by other researchers, thus ensuring the reproducibility of the research results and facilitating future enhancements.
As compared to a recent work in~\cite{s19051196}, where a machine learning based EESM is proposed for NR systems, this paper considers a complete NR PHY abstraction including the details on SINR-BLER curves under various NR settings, provides support for multiple MCS tables, and includes up to 256-QAM (the work in \cite{s19051196} supports up to 64-QAM). Also, as compared other L2SM works in the literature, we do not only present the PHY abstraction model but also integrate it into an NR based SLS and present ns-3 based end-to-end simulation results to illustrate its usage. 

This paper is organized as follows. In Section~\ref{sec:L2SM},
the EESM method used in the paper is presented.
In Section~\ref{sec:PHYabstraction}, details of the PHY abstraction model with the description of its every block and the related NR features affecting the model are provided. Section~\ref{sec:eval} includes results using end-to-end simulations in ns-3. 
Conclusions are drawn in Section~\ref{sec:conc}.

\section{Exponential Effective SINR Mapping}
\label{sec:L2SM}
In this section, we provide details on the EESM based SINR compression techniques considered in this paper to derive the effective SINR based on the SINR values per RB~\cite{brueninghaus:05}, for both single transmission and when combining retransmissions using Hybrid Automatic Repeat reQuest (HARQ).

\subsection{Effective SINR for single transmission}
\label{sec:nonharq}
In case of EESM~\cite{brueninghaus:05}, the  mapping
function  is   exponential and the effective SINR for single transmission is obtained as:
\begin{equation}
    \text{SINR}_{\text{eff}} = {-}\beta \ln \Big( \frac{1}{|\upsilon|}\sum_{n \in \upsilon} \exp\big({-}\frac{\text{SINR}_n}{\beta}\big)\Big), \label{sinr}
\end{equation}
where $\text{SINR}_n$ is the SINR value in the $n$th RB, $\upsilon$ is the set of allocated RBs, and $\beta$ is a parameter that needs to be optimized (as mentioned in the introduction).
The details on $\beta$ calibration are given in Section~\ref{sec:betaopt}.

Eq.~\eqref{sinr} holds true when the same MCS is used over all RBs of the transmission, which is a common assumption in 4G and 5G systems. 

\subsection{Effective SINR for combined retransmissions}
\label{sec:harq}
The NR scheduler works on a slot basis and has a dynamic nature~\cite{TS38300}. For example, it may assign different sets of OFDM symbols in time and RBs in frequency for transmissions and the corresponding redundancy versions. However, it always assigns an integer multiple of the RB consisting of 12 resource elements in frequency domain and 1 OFDM symbol in time domain.
In this work, for simplicity, we assume that retransmissions (including the first transmission and the corresponding redundancy versions) of the same HARQ process use the same MCS and the same number of RBs, although the specific RBs' time/frequency positions within a slot may vary in between the retransmissions. Also, we consider that the SINRs experienced on each RB may vary through retransmissions. 

The two HARQ methods used in communications standards are Chase Combining (HARQ-CC) and Incremental Redundancy (HARQ-IR). The EESM for combined retransmissions varies with the underline HARQ method~\cite{80216m}. 

\textbf{HARQ-CC:} In HARQ-CC, every retransmission contains the same coded bits (information and coding bits). Therefore, the effective code rate (ECR) after the $q$th retransmission remains the same as after the first transmission. In this case, the SINR values of the corresponding resources are summed across the retransmissions, and the combined SINR values are used for EESM~\cite{80216m}. After $q$ retransmissions, the effective SINR using EESM
is computed as:
\begin{equation}
    \text{SINR}_{\text{eff}} = {-}\beta \ln \Big( \frac{1}{|\omega|}\sum_{m \in \omega} \exp \big({-}\frac{1}{\beta}\sum_{j=1}^q \text{SINR}_{m,j}\big)\Big), \label{sinr_cc}
\end{equation}
where SINR$_{m,j}$ is the SINR experienced by the $m$th RB in the $j$th retransmission, and $\omega$ is the set of RBs to be combined.

\begin{figure*}[!t]
	\centering
	\includegraphics[width=0.75\textwidth]{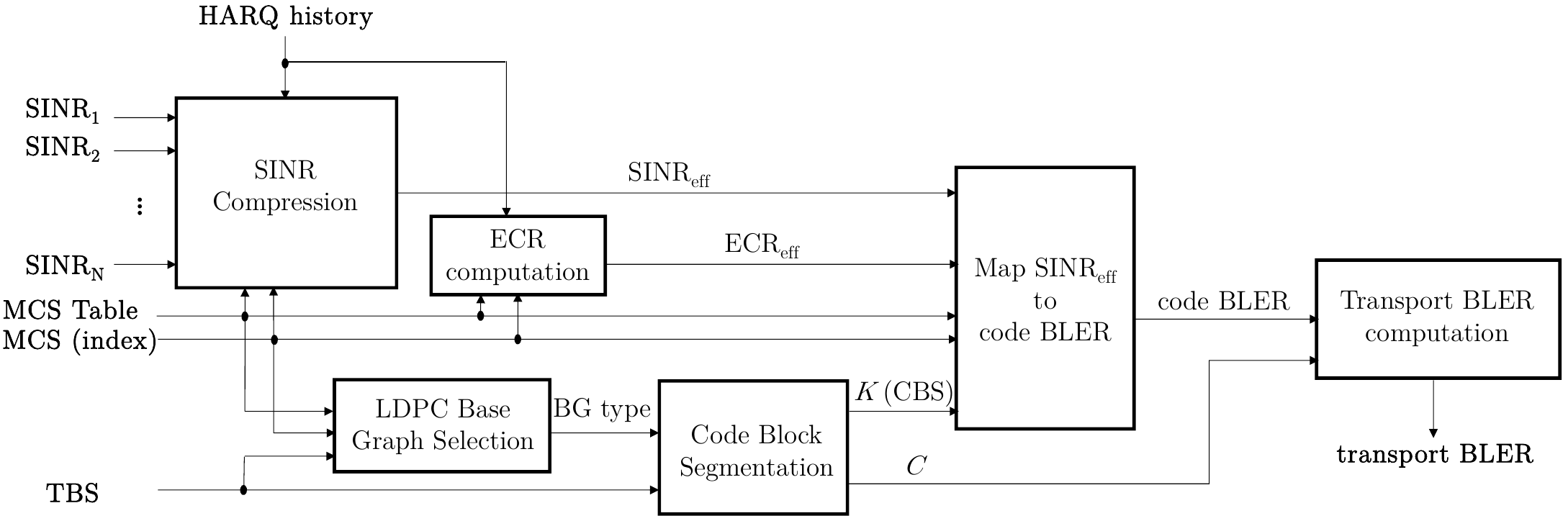}
	\captionsetup{justification=centering, font=small}
	\caption{NR PHY abstraction model.}
	\label{fig_comp}
	\vspace{-0.4cm}
\end{figure*}

\textbf{HARQ-IR:} In HARQ-IR, every retransmission contains different coded bits than the previous one. The different retransmissions typically use a different set of coding bits~\cite{80216m}. Therefore, both the effective SINR and the ECR need to be recomputed after each retransmission.
The ECR after $q$ retransmissions is obtained as:
\begin{equation}
    \text{ECR}_{\text{eff}} = \frac{X}{\sum_{j=1}^q C_j}, \label{ecr}
\end{equation}
where $X$ is the number of information bits and $C_j$ is the number of coded bits in the $j$th retransmission. 
The effective SINR using EESM after $q$ retransmissions is given by:
\begin{equation}
    \text{SINR}_{\text{eff}} = {-}\beta \ln \Big( \frac{1}{q|\omega|}\sum_{j=1}^q\sum_{m \in \omega}\exp\big({-}\frac{ \text{SINR}_{m,j}}{\beta}\big)\Big), \label{sinr_ir}
\end{equation}
where 
SINR$_{m,q}$ is the SINR experienced by the $m$th RB in the $q$th retransmission, and $\omega$ is the set of RBs to be combined, which is maintained over retransmissions.

\section{NR PHY Abstraction}
\label{sec:PHYabstraction}
The overall NR PHY abstraction model that we propose for SLS is shown in Fig.~\ref{fig_comp}. The L2SM process receives inputs consisting of a vector of SINRs per allocated RB, the MCS selection (including MCS index and the MCS table to which it refers), the TBS delivered to PHY, and the HARQ history. Then, 
it provides as output the BLER of the MAC transport block. The  model consists of the following blocks: SINR compression, LDPC base graph (BG) selection, segmentation of a transport block into one or multiple code blocks (known as code block segmentation), mapping of the effective SINR to BLER for each PHY code block (denoted as \textit{code BLER} for short), and mapping of code BLERs to the transport BLER.

The HARQ history depends on the HARQ method. In HARQ-CC, the HARQ history contains the SINR per allocated RB, whereas for HARQ-IR, the HARQ history contains the last computed effective SINR and number of coded bits of each of the previous retransmissions. Given the SINR vector and the HARQ history, the effective SINR is computed according to EESM, as detailed in Section~\ref{sec:L2SM}. The optimization of EESM, i.e., finding an optimal value of $\beta$, is performed using a NR-compliant LLS as described in Section~\ref{sec:betaopt}. The details of LDPC BG selection, which uses TBS and MCS selection, are provided in~Section~\ref{sec:mcs}. Once the BG selection is known, the code block segmentation (if needed) is performed to derive the number of code blocks and the number of bits in each code block, which is also known as code block size (CBS). The details on the code block segmentation procedure can be found in Section~\ref{sec:mcs}. Given the effective SINR, the ECR (as described in Section~\ref{sec:L2SM}), the MCS selection, and the CBS, the corresponding code BLER can be found as described in Section~\ref{sec:tables} using SINR-BLER lookup tables obtained from NR-compliant LLS.  
Finally, based on the number of code blocks and the code BLER, the transport BLER of the transport block is obtained as detailed in Section~\ref{sec:tbler}.

\subsection{MCS selection, BG selection, and code block segmentation}
\label{sec:mcs}
\textbf{MCS}: NR defines three tables of MCSs: MCS Table1 (up to 64-QAM), MCS Table2 (up to 256-QAM), and MCS Table3 (up to 64-QAM with low spectral efficiency), which are given by Tables 5.1.3.1-1 to 5.1.3.1-3 in \cite{TS38214}.  
A base station (known as gNB in NR) can indicate the table selection to a user (UE) either semi-statically or dynamically, and the MCS index selection is communicated to the UE for each transmission.
Each MCS index is quantized by 5 bits and defines an ECR, a modulation order, and the resulting spectral efficiency (SE). 
In this work, we focus on MCS Table1 and MCS Table2. The MCS Table1 includes from MCS0 (ECR=0.12, QPSK, SE=0.23 bits/s/Hz) to MCS28 (ECR=0.94, 64-QAM, SE=5.55 bits/s/Hz), whereas the MCS Table2 has MCS indices from MCS0 (ECR=0.12, QPSK, SE=0.23 bits/s/Hz) to MCS27 (ECR=0.93, 256-QAM, SE=7.40 bits/s/Hz).
As shown in Fig.~\ref{fig_comp}, the MCS Table (1 or 2) and the MCS index (0 to 28 for MCS Table1, and 0 to 27 for MCS Table2) are inputs for the NR PHY abstraction. 

\textbf{MCS selection}: MCS selection in NR is an implementation specific procedure. However, NR defines the Channel Quality Indicator (CQI), which is reported by the UE and can be used for MCS index selection at the gNB. 
NR defines three tables of 4-bit CQIs (see Tables 5.2.2.1-1 to 5.2.2.1-3 in~\cite{TS38214}), each table being associated with one MCS table. 

The PHY abstraction model shown in Fig.~\ref{fig_comp} is used as an error model, but it can also be used for link adaptation~\cite{mezzavilla:12}, i.e., to determine an MCS that satisfies a target transport BLER (e.g., 10$\%$) based on the actual channel conditions. In that case, for a given set of SINR values, a target transport BLER, an MCS table, and considering a transport block composed of the group of RBs in the band (termed the CSI reference resource~\cite{TS38214}), the highest MCS index that meets the target transport BLER constraint is selected at the UE. Such value is then reported through the associated CQI index to the gNB. 

\textbf{LDPC BG selection}: BG selection in NR is based on the following conditions~\cite[Sect. 6.2.2]{TS38212}. Assuming $R$ as the ECR of the selected MCS and $A$ as the TBS (in bits), then, 
\begin{itemize}
    \item LDPC base graph 2 (BG2) is selected if $A{\le} 292$ with any value of $R$, or if $R{\le} 0.25$ with any value of $A$, or if $A{\le} 3824$ with $R{\le} 0.67$, 
    \item otherwise, the LDPC base graph 1 (BG1) is selected.
\end{itemize}

\textbf{Code block segmentation}: 
Code block segmentation for LDPC coding in NR occurs when the number of total bits in a transport block including cyclic redundancy check (CRC) is larger than the maximum CBS ($K_{\text{cb}}$), where, $K_{\text{cb}}{=}8448$ bits for LDPC BG1 and $K_{\text{cb}}{=}3840$ bits for LDPC BG2. If code block segmentation occurs, each transport block is split into $C$ code blocks of $K$ bits each, and for each code block, an additional CRC sequence of $L{=}24$ bits is appended to recover the segmentation during the decoding process. The segmentation process takes LDPC BG selection and LDPC lifting size into account, the complete details of which can be found in~\cite[Sect. 5.2.2]{TS38212}.

\subsection{SINR Compression ($\beta$ optimization)}
\label{sec:betaopt}
In EESM, given an experimental BLER measured in a fading channel with a specific MCS, the value of $\beta$ is calibrated such that the effective SINR of that channel approximates to the SINR that would produce the same BLER, with the same MCS, in AWGN channel conditions~\cite{cipriano:08}. 

In order to obtain the optimal values of $\beta$, we implement a NR-compliant LLS and use a calibration technique described in~\cite{cipriano:08}. The simulation parameters are shown in Table~\ref{table:lls}. We use tapped delay line (TDL) based fading channel models recommended by 3GPP in~\cite{TR38901}. A collection of LOS (TDL-D) and NLOS (TDL-A) channel models ranging in delay spread from 30 ns to 316 ns are used.
For NR, subcarrier spacing (SCS) of 30 kHz and 60 kHz are simulated.

For each MCS, an optimal value of $\beta$ is derived as the argument that minimizes the following error function:
\begin{equation}
\begin{aligned}
       \beta_{\text{opt}} = \operatorname*{argmin}_\beta \frac{1}{|\zeta||\eta|} & \sum\limits_{k \in \eta}\sum\limits_{l \in \zeta} \Big|\text{log}_{10}(\text{BLER}_p(\mathbf{H}_l,\sigma_k^2)) - \\
      & \text{log}_{10}\text{(BLER}_r(\text{SINR}_\text{eff}(\mathbf{H}_l,\sigma_k^2, \beta)))\Big|^2,
\end{aligned}
\end{equation}
where $\mathbf{H}_l$ represents a channel realization, and $\sigma_k^2$ denotes an AWGN variance realization. The terms $\text{BLER}_p(\mathbf{H}_l,\sigma_k^2)$ and $\text{BLER}_r(\text{SINR}_\text{eff}(\mathbf{H}_l,\sigma_k^2, \beta))$ represent the measured BLER and the BLER on AWGN channel for the $\text{SINR}_\text{eff}$ obtained using EESM, respectively. The calibration is done over all the different channel realizations (represented by the set $\zeta$) of the chosen channel models mentioned previously and different values of noise variances (represented by the set $\eta$). 

\begin{table}
\centering
\footnotesize
\begin{tabular}{|m{2cm}||m{2.6cm}|m{2.6cm}|} 
 \hline
   & \textbf{Indoor Simulation} & \textbf{Outdoor simulation} \\
 \hline
 \hline
  scenario & Indoor office & UMi Street-canyon \\
 \hline
  channel model & TDL-A, TDL-D &  TDL-A, TDL-D \\
 \hline
   frequency & \multicolumn{2}{c|}{5 GHz}\\
 \hline
   bandwidth & \multicolumn{2}{c|}{20 MHz} \\
 \hline
   SCS  & \multicolumn{2}{c|}{30 kHz, 60 kHz} \\
   \hline
   delay spread & 30 ns, 53 ns & 93 ns, 316 ns \\
 \hline
 PDSCH config & \multicolumn{2}{c}{12 OFDM symbols, no PTRS, 1 DMRS symbol} \\
 \hline
\end{tabular}
\caption{\small Channel assumptions in LLS for $\beta$ optimization.}
\label{table:lls}
\vspace{-0.4cm}
\end{table}

Using this calibration technique, we obtain the optimal $\beta$ values, which are shown in Table~\ref{tab_beta} for each MCS index in MCS Table1 and MCS Table2.

\begin{remark}
From our simulation results it is observed that the effective SINR mapping is insensitive to the SCS (or, in other words, to the NR numerology), in addition to the carrier frequency. 
Therefore, the proposed model \textit{is valid for all NR frequency ranges and numerologies}.
\end{remark}

\begin{table}
\centering
\footnotesize
\begin{tabular}{|m{0.7cm}|m{0.9cm}|m{0.9cm}|| m{0.7cm}|m{0.9cm}|m{0.9cm}|} 
 \hline
  \multirow{2}{*} {\textbf{MCS}} & \multicolumn{2}{|l|}{\quad \quad optimal $\beta$} & \multirow{2}{*} {\textbf{MCS}} & \multicolumn{2}{|l|}{\quad \quad optimal $\beta$}\\
  & Table1 & Table2 &  & Table1 & Table2\\
 \hline
 \hline
  0 & 1.60 & 1.60 & 15 & 6.16 & 19.33 \\
 \hline
  1 & 1.61 & 1.63 & 16 & 6.50 & 21.85 \\
 \hline
   2 & 1.63 & 1.67 & 17 & 9.95 & 24.51 \\
 \hline
   3 & 1.65 & 1.73 & 18 & 10.97 & 27.14 \\
 \hline
   4 & 1.67 & 1.79 & 19 & 12.92 & 29.94 \\  
 \hline
  5 & 1.70 & 4.27 & 20 & 14.96 & 56.48 \\
 \hline
  6 & 1.73 & 4.71 & 21 & 17.06 & 65.00 \\
 \hline
  7 & 1.76 & 5.16 & 22 & 19.33 & 78.58 \\
 \hline
  8 & 1.79 & 5.66 & 23 & 21.85 & 92.48 \\
 \hline
  9 & 1.82 & 6.16 & 24 & 24.51 & 106.27 \\
  \hline
  10 & 3.97 & 6.50 & 25 & 27.14 & 118.74 \\
 \hline
  11 & 4.27 & 10.97 & 26 & 29.94 & 126.36 \\
 \hline
  12 & 4.71 & 12.92 & 27 & 32.05 & 132.54 \\
 \hline
   13 & 5.16 & 14.96 & 28 & 34.28 & - \\
 \hline
   14 & 5.66 & 17.06 & &  & \\
 \hline
 \end{tabular}
\caption{\small Optimal $\beta$ values for each MCS.}
\label{tab_beta}
\vspace{-0.4cm}
\end{table}

\begin{figure}[!t]
\vspace{-0.2cm}
	\centering
	\includegraphics[width=0.46\textwidth]{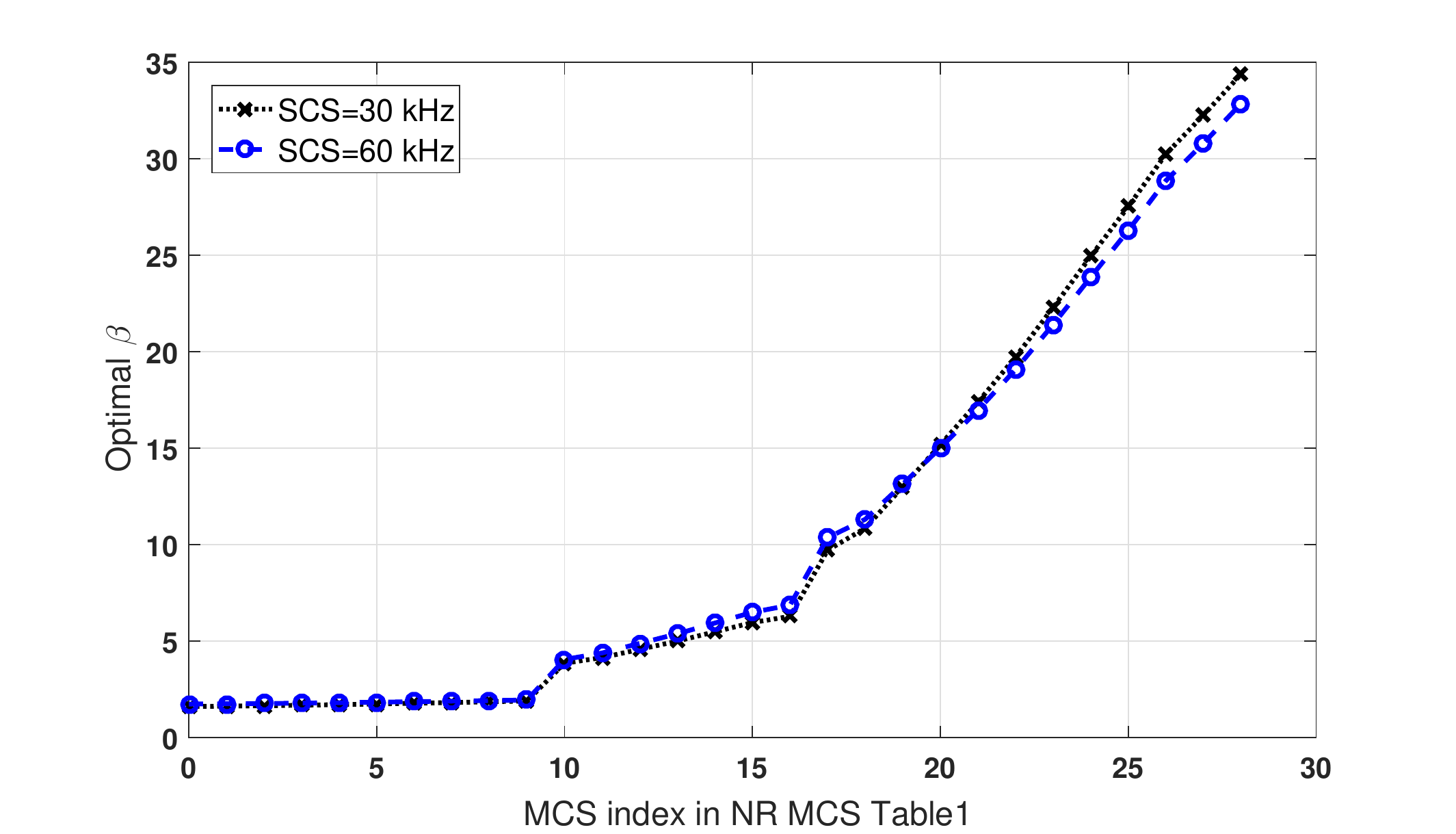}
	\captionsetup{justification=centering, font=small}
	\caption{Optimal $\beta$ values for each MCS in MCS Table1, when using 30 kHz and 60 kHz SCS.
	}
	\label{fig_scs}
	\vspace{-0.4cm}
\end{figure}

To illustrate such conclusion, in Fig.~\ref{fig_scs} we show the optimal $\beta$ values when using 30 kHz SCS and 60 kHz SCS, separately, in the calibration technique. There is no clear trend emerging to indicate that the optimal $\beta$ depends on the SCS. Such result is used just to illustrate the remark, while optimal $\beta$ values in Table~\ref{tab_beta} do consider the results averaged over both SCSs.

\subsection{Effective SINR to code BLER mapping}\label{sec:tables}
The transport BLER depends on SINR, MCS, and resource allocation. In SLS, once we have the effective SINR (using the optimal $\beta$ described in Section~\ref{sec:betaopt}) for the given MCS, resource allocation, and channel model, we need SINR-BLER lookup tables to find the corresponding transport BLER. 

In order to obtain SINR-BLER mappings, we perform extensive simulations using our NR-compliant LLS. For these simulations, we consider 28 GHz carrier frequency, SCS of 120 kHz, and bandwidth of 200 MHz. For each MCS (both in MCS Table1 and Table2), various resource allocation (with varying number of RBs from 1 to 132 and varying number of OFDM symbols from 1 to 10) are simulated. Given the resource allocation, the corresponding value of TBS, CBS, number of code blocks, LDPC BG selection, and LDPC lifting size can be derived. In our LLS, the TBS remains below the maximum CBS (i.e., 8448 bits  for  LDPC  BG1 or 3840 bits  for  LDPC  BG2), therefore, there is no need of code block segmentation in LLS, instead, code block segmentation is integrated into the proposed NR PHY abstraction model to speed up the simulation rate (see Fig.~\ref{fig_comp}). With this setting, we run our LLS and generate code BLER vs SINR curves for each MCS with different values of CBSs (with the corresponding BG selection).

\begin{figure}[!t]
\vspace{-0.3cm}
	\centering
	\includegraphics[width=0.47\textwidth]{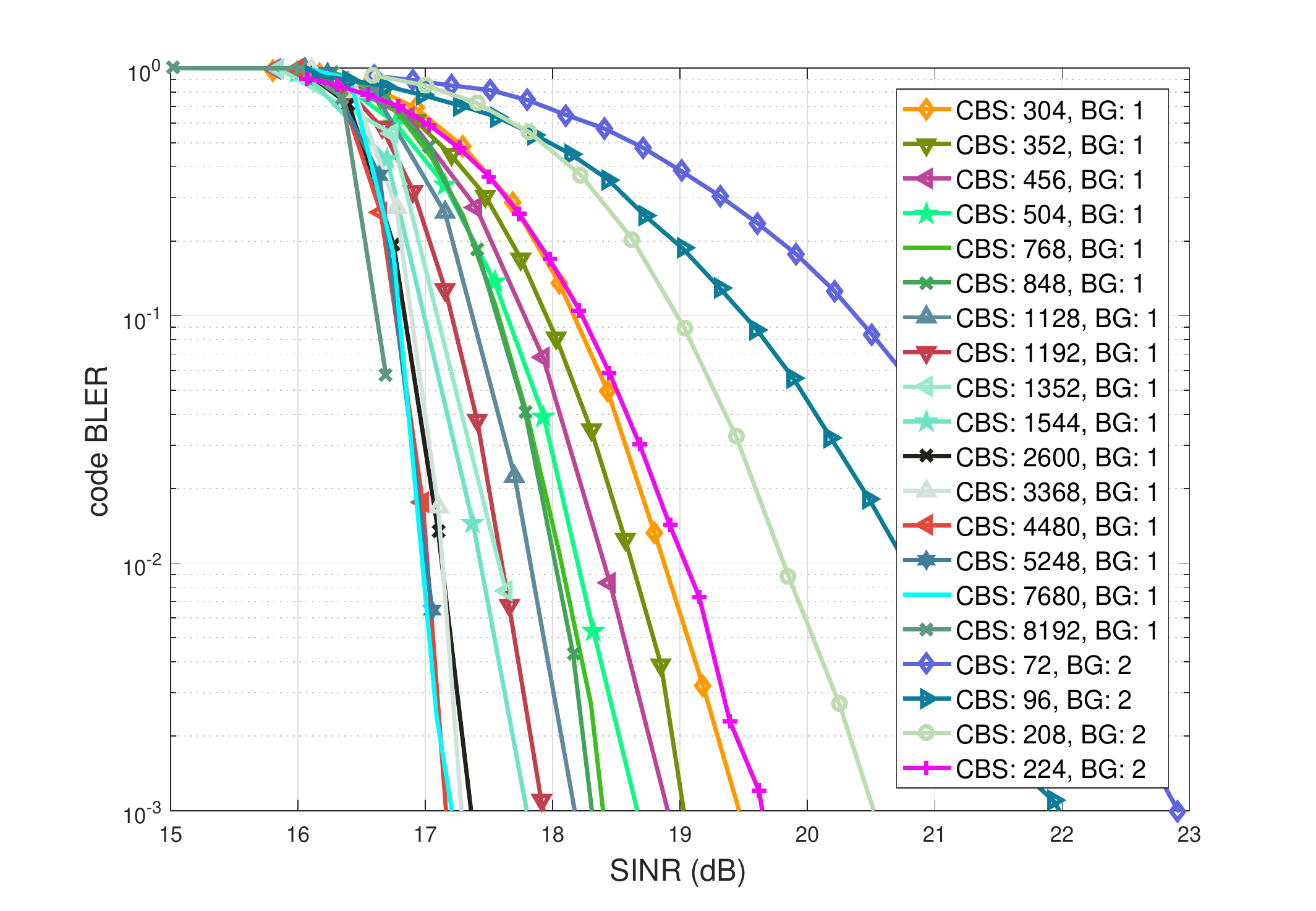}
	\captionsetup{justification=centering, font=small}
	\caption{code BLER vs SINR (dB) for 
	MCS23 of MCS Table1.
	}
	\label{fig_bler1}
	\vspace{-0.5cm}
\end{figure}
\begin{figure}[!t]
	\centering
	\includegraphics[width=0.47\textwidth]{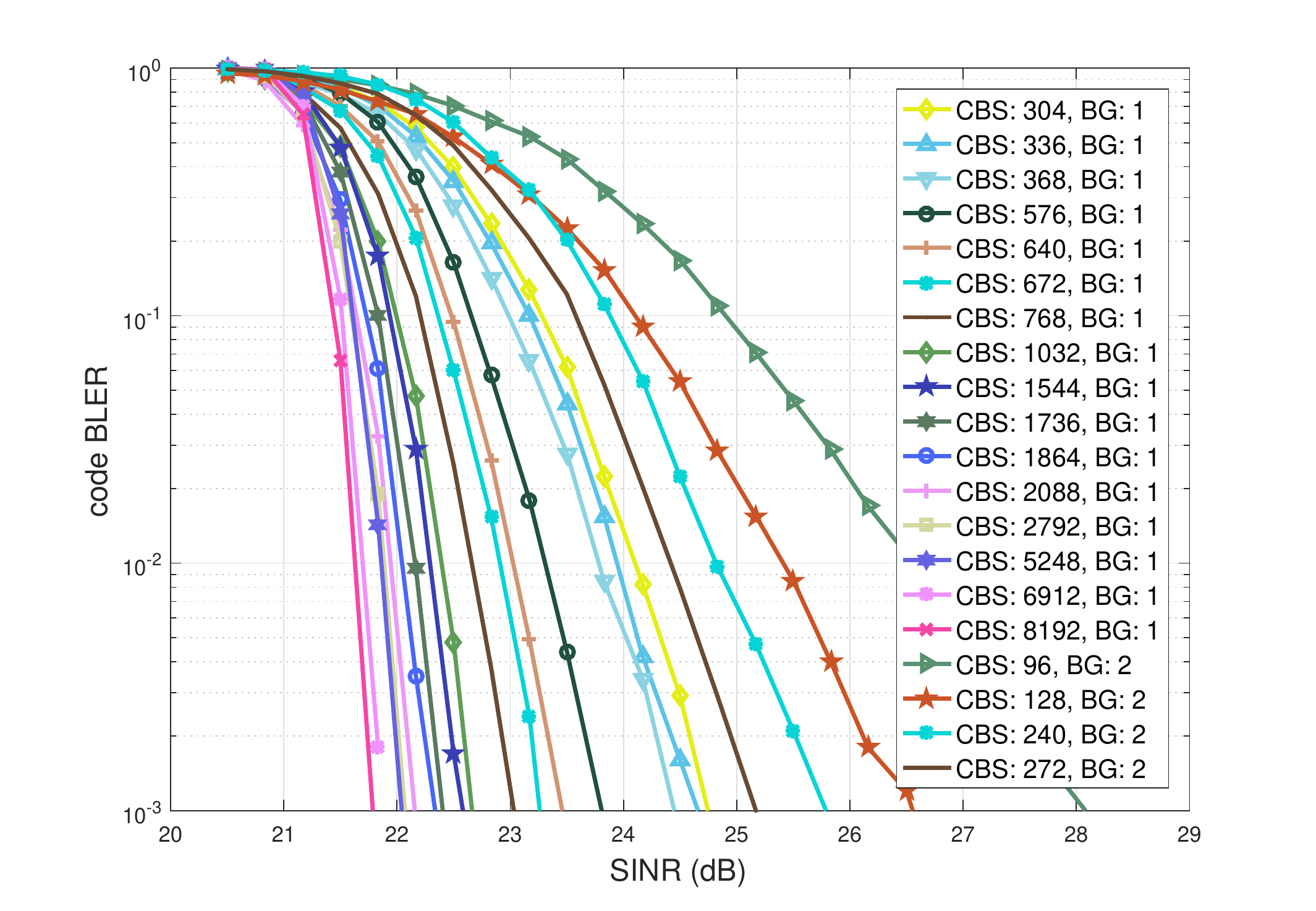}
	\captionsetup{justification=centering, font=small}
	\caption{code BLER vs SINR (dB) for MCS21 of MCS Table2.
	}
	\label{fig_bler2}
	\vspace{-0.4cm}
\end{figure}

Due to space constraints we cannot include all the SINR-code BLER curves in this paper, instead, we only show the results for two specific MCS indices, i.e., MCS23 of MCS Table1 in Fig.~\ref{fig_bler1} and MCS21 of MCS  Table2 in Fig.~\ref{fig_bler2}. The data for all the curves can be found in the ns-3 based NR SLS~\cite{5GLENA} (https://5g-lena.cttc.es).
For each case, different CBSs are simulated, and for each CBS, the selected BG type is indicated in the legend. 
Note that because MCS23 of MCS Table1 is equivalent to MCS16 of MCS Table2, therefore, the range of SINRs in Fig.~\ref{fig_bler1} is lower than the SINR range of Fig.~\ref{fig_bler2}. As it can be seen from the plots, the CBS has a high impact on the actual BLER performance for a given MCS, which was observed also in LTE~\cite{mezzavilla:12}. As the CBS increases, it provides better code BLER performance for a fixed SINR. 

Note that the SINR-code BLER curves obtained from LLS are quantized and consider a subset of CBSs. Accordingly, in the SLS, we implement a worst case approach to determine the code BLER value by using lower bounds of the actual CBS and effective SINR. In the PHY abstraction for HARQ-IR, for simplicity and according to the obtained curves, we limit the effective ECR in Eq.~\eqref{ecr} by the lowest ECR of the MCSs that have the same modulation order as the selected MCS index.

\subsection{Transport BLER computation}
\label{sec:tbler}
As mentioned in the previous section, each baseline SINR-BLER curve was estimated without code block segmentation in the LLS, ensuring that the transport BLER was equal to the code BLER.
However, in the SLS, we simulate TBSs that may require code block segmentation. Therefore, there is a need to convert the code BLER found from the LLS's lookup table to the transport BLER for the given TBS. 

The code BLERs of the $C$ code blocks (as determined by the code block segmentation described in Section~\ref{sec:mcs}) are combined to get the BLER of a transport block as:
\begin{equation}
    \text{TBLER} = 1- \prod_{i=1}^{C} (1-\text{CBLER}_i) \approxeq 1- (1-\text{CBLER})^{C},
\end{equation}
where the last approximate equality holds because code block segmentation in NR generates code blocks of roughly equal sizes\footnote{NR allows rate matching to vary code block sizes slightly to assure code blocks align on OFDM symbol boundaries.}.

\section{End-to-End Evaluation in ns-3}
\label{sec:eval}
As previously mentioned, the PHY abstraction model proposed in Section~\ref{sec:PHYabstraction} has been integrated into an NR based SLS, particularly in the NR SLS of the popular and open-source ns-3~\cite{5GLENA}, thus giving life and promoting the use of the proposed model. In this section, an end-to-end evaluation to illustrate the usability of the model is performed. 
Different sets of NR configurations are used, including fixed MCS, adaptive MCS, different HARQ methods (HARQ-CC and HARQ-IR), and the two NR MCS tables (MCS Table1 and Table2).

\subsection{Simulation Scenario}
We consider a gNB-UE link in an Urban Micro scenario with 28 GHz carrier frequency, SCS of 120 kHz, and 100 MHz channel bandwidth. For the simulations, the gNB-UE 2D distance is varied (within values 10m, 30m, 50m, 70m) to emulate different channel qualities. The heights of the gNB and the UE are set to 10 m and 1.5 m, respectively. The fast fading based 3GPP channel model~\cite{TR38901} is used. 
Uniform planar arrays are used for both the gNB and the UE. The number of antennas is set to 4$\times$8 at the gNB and 2$\times$4 at the UE, with 4 dBm transmit power at the gNB. Noise power spectral density of -174 dBm/Hz and noise figure of 5 dB are considered. Only downlink UDP (User Datagram Protocol) traffic is simulated, in which, one packet (with size of 100 Bytes) every 200 ms is sent over a period of 50 s. The Radio Link Control (RLC) is used in unacknowledged mode (UM).
Two slots of PHY-MAC processing delay (i.e., 250 us) and 100 us of decoding latency are considered. We update the channel every 150 ms so that every packet encounters a different channel realization, thus getting statistical significance with respect to the fading. 

\begin{table*}[t!]
\centering
\footnotesize
\begin{tabular}{|m{3.2cm}|| m{0.9cm}|m{0.5cm}|m{0.5cm}||m{0.9cm}|m{0.5cm}|m{0.5cm} ||m{0.9cm}|m{0.5cm}|m{0.5cm}||m{0.9cm}|m{0.5cm}|m{0.5cm}|} 
 \hline
   gNB-UE distance:  & \multicolumn{3}{c||}{10m} & \multicolumn{3}{c||}{30m} & \multicolumn{3}{c||}{50m} & \multicolumn{3}{c|}{70m} \\
 \hline
  performance metrics: & delay & APP loss & PHY loss & delay & APP loss & PHY loss & delay & APP loss & PHY loss & delay & APP loss & PHY loss\\
 \hline
 \hline
  Table1 MCS13 HARQ-CC & 0.49ms & 0$\%$  & 0$\%$ & 0.90ms & 0$\%$  & 35$\%$ & 12.49ms & 97$\%$  & 99$\%$ & - & 100$\%$  & 100$\%$\\
  Table1 MCS13 HARQ-IR & 0.49ms & 0$\%$  & 0$\%$ & 0.90ms & 0$\%$  & 35$\%$ & 5.85ms & 35$\%$  & 82$\%$ & - & 100$\%$  & 100$\%$\\
  Table2 MCS7 HARQ-CC & 0.49ms & 0$\%$  & 0$\%$ & 0.90ms & 0$\%$  & 35$\%$ & 12.49ms & 97$\%$  & 99$\%$ & - & 100$\%$  & 100$\%$\\
  Table2 MCS7 HARQ-IR & 0.49ms & 0$\%$  & 0$\%$ & 0.90ms & 0$\%$  & 35$\%$ & 8.29ms & 54$\%$  & 88$\%$ & - & 100$\%$  & 100$\%$\\
 \hline
 \end{tabular}
\caption{\small End-to-end evaluation with fixed MCS, using different MCS Tables but the same modulation order and ECR.}
\label{tab_fixmcs1}
\vspace{-0.1cm}
\end{table*}

\begin{table*}[t!]
\centering
\footnotesize
\begin{tabular}{|m{3.2cm}|| m{0.9cm}|m{0.5cm}|m{0.5cm}||m{0.9cm}|m{0.5cm}|m{0.5cm} ||m{0.9cm}|m{0.5cm}|m{0.5cm}||m{0.9cm}|m{0.5cm}|m{0.5cm}|}  
 \hline
   gNB-UE distance:  & \multicolumn{3}{c||}{10m} & \multicolumn{3}{c||}{30m} & \multicolumn{3}{c||}{50m} & \multicolumn{3}{c|}{70m} \\
 \hline
  performance metrics: & delay & APP loss & PHY loss & delay & APP loss & PHY loss & delay & APP loss & PHY loss & delay & APP loss & PHY loss\\
 \hline
 \hline
  Table2 MCS7 HARQ-CC & 0.49ms & 0$\%$  & 0$\%$ & 0.90ms & 0$\%$  & 35$\%$ & 12.49ms & 97$\%$  & 99$\%$ & - & 100$\%$  & 100$\%$\\
  Table2 MCS11 HARQ-CC & 0.49ms & 0$\%$  & 0$\%$ & 2.07ms & 0$\%$  & 65$\%$ & - & 100$\%$  & 100$\%$ & - & 100$\%$  & 100$\%$\\
  Table2 MCS7 HARQ-IR & 0.49ms & 0$\%$  & 0$\%$ & 0.90ms & 0$\%$  & 35$\%$ & 8.29ms & 54$\%$  & 88$\%$ & - & 100$\%$  & 100$\%$\\
  Table2 MCS11 HARQ-IR & 0.49ms & 0$\%$  & 0$\%$ & 1.71ms & 0$\%$  & 62$\%$ & - & 100$\%$  & 100$\%$ & - & 100$\%$  & 100$\%$\\
 \hline
 \end{tabular}
\caption{\small End-to-end evaluation with fixed MCS, using different MCS indices of the same MCS Table (Table2).}
\label{tab_fixmcs2}
\vspace{-0.4cm}
\end{table*}

For the performance metrics, the end-to-end delay of UDP packets (`delay' in ms), the packet loss at application layer (`APP loss' in $\%$), and transmission data failures at PHY layer (`PHY loss' in $\%$) are collected.
Since higher packet losses are expected with increasing the gNB-UE distance, to alleviate the impact of the RLC UM timers on the end-to-end delay, the 
RLC UM is configured with a reordering window timer of 10 ms and a reporting buffer status timer from RLC to MAC of 1 ms. 
In the following sections, we first discuss the results with fixed MCS in Section \ref{sec:res1} and then with adaptive MCS in Section \ref{sec:res2}.

\begin{table*}[t!]
\centering
\footnotesize
\begin{tabular}{|m{2.4cm}|| m{0.8cm}|m{0.4cm}|m{0.4cm}|m{0.35cm}||m{0.8cm}|m{0.4cm}|m{0.4cm}|m{0.35cm} ||m{0.8cm}|m{0.4cm}|m{0.4cm}|m{0.35cm}||m{0.8cm}|m{0.4cm}|m{0.4cm}|m{0.35cm}|}  
 \hline
   gNB-UE distance:  & \multicolumn{4}{c||}{10m} & \multicolumn{4}{c||}{30m} & \multicolumn{4}{c||}{50m} & \multicolumn{4}{c|}{70m} \\
 \hline
  performance metrics: & delay & app loss & phy loss & mcs & delay & app loss & phy loss & mcs & delay & app loss & phy loss & mcs & delay & app loss & phy loss & mcs \\
 \hline
 \hline
  Table1 HARQ-CC & 0.49ms & 0$\%$ & 0$\%$ & 27 & 0.49ms & 0$\%$ & 0$\%$ & 11 & 0.53ms & 0$\%$ & 0$\%$ & 0 & 1.99ms & 0$\%$ & 66$\%$ & 0 \\
  Table1 HARQ-IR & 0.49ms & 0$\%$ & 0$\%$ & 27 & 0.49ms & 0$\%$ & 0$\%$ & 11 & 0.53ms & 0$\%$ & 0$\%$ & 0 & 1.47ms & 0$\%$ & 55$\%$ & 0 \\
  Table2 HARQ-CC & 0.53ms & 0$\%$ & 4$\%$ & 25 & 0.49ms & 0$\%$ & 0$\%$ & 5 & 0.53ms & 0$\%$ & 0$\%$ & 0 & 1.99ms & 0$\%$ & 66$\%$ & 0 \\
  Table2 HARQ-IR & 0.53ms & 0$\%$ & 4$\%$ & 25 & 0.49ms & 0$\%$ & 0$\%$ & 5 & 0.53ms & 0$\%$ & 0$\%$ & 0 & 1.47ms & 0$\%$ & 55$\%$ & 0 \\
 \hline
 \end{tabular}
\caption{\small End-to-end evaluation with link adaptation based on the error model, using different MCS Tables and HARQ methods.}
\label{tab_amc1}
\vspace{-0.1cm}
\end{table*}

\begin{table*}[t!]
\centering
\footnotesize
\begin{tabular}{|m{2.4cm}|| m{0.8cm}|m{0.4cm}|m{0.4cm}|m{0.35cm}||m{0.8cm}|m{0.4cm}|m{0.4cm}|m{0.35cm} ||m{0.8cm}|m{0.4cm}|m{0.4cm}|m{0.35cm}||m{0.8cm}|m{0.4cm}|m{0.4cm}|m{0.35cm}|}  
 \hline
  gNB-UE distance:  & \multicolumn{4}{c||}{10m} & \multicolumn{4}{c||}{30m} & \multicolumn{4}{c||}{50m} & \multicolumn{4}{c|}{70m} \\
 \hline
  performance metrics: & delay & app loss & phy loss & mcs & delay & app loss & phy loss & mcs & delay & app loss & phy loss & mcs & delay & app loss & phy loss & mcs \\
 \hline
 \hline
  Table1 HARQ-CC & 0.49ms & 0$\%$ & 0$\%$ & 27 & 0.50ms & 0$\%$ & 0$\%$ & 6 & 0.53ms & 0$\%$ & 0$\%$ & 0 & 1.99ms & 0$\%$ & 66$\%$ & 0 \\  Table1 HARQ-IR & 0.49ms & 0$\%$ & 0$\%$ & 27 & 0.50ms & 0$\%$ & 0$\%$ & 6 & 0.53ms & 0$\%$ & 0$\%$ & 0 & 1.47ms & 0$\%$ & 55$\%$ & 0 \\ 
  Table2 HARQ-CC & 0.49ms & 0$\%$ & 0$\%$ & 23 & 0.50ms & 0$\%$ & 0$\%$ & 3 & 0.53ms & 0$\%$ & 0$\%$ & 0 & 1.99ms & 0$\%$ & 66$\%$ & 0 \\ 
  Table2 HARQ-IR & 0.49ms & 0$\%$ & 0$\%$ & 23 & 0.50ms & 0$\%$ & 0$\%$ & 3 & 0.53ms & 0$\%$ & 0$\%$ & 0 & 1.47ms & 0$\%$ & 55$\%$ & 0 \\ 
 \hline
 \end{tabular}
\caption{\small End-to-end evaluation with link adaptation based on the Shannon bound, using different MCS Tables and HARQ methods.}
\label{tab_amc2}
\vspace{-0.4cm}
\end{table*}

\subsection{Results with fixed MCS}
\label{sec:res1}
The first simulation campaign assesses the performance with fixed MCS.
In Table~\ref{tab_fixmcs1}, we show the results for different NR MCS tables (Table1 and Table2) when using the MCS indices that correspond to the same modulation order and ECR, i.e., MCS13 of MCS Table1 and MCS7 of MCS Table2. Then, in Table~\ref{tab_fixmcs2}, we show the results for different MCS indices (MCS7 and MCS11) of the same NR MCS Table (Table2). 

For the same modulation order and ECR (Table~\ref{tab_fixmcs1}), we observe the following. Firstly, for good and medium channel qualities (10m and 30m gNB-UE distances), the same delay performance is obtained in all cases and no losses occur at application layer, although HARQ retransmission combining is needed at 30m distance.
In the case of 50m distance, packet losses at application layer arise, therefore, the RLC-UM starts to play a role and end-to-end packet delays are affected by the reordering window timer of 10 ms.
For such distance (i.e., 50m), results are better with HARQ-IR as compared to HARQ-CC both in terms of delay and application throughput. This is because in HARQ-IR, the equivalent ECR is reduced when retransmissions are combined, which provides more robustness to channel impairments. 

Secondly, as expected, the results for Table1 MCS13 and Table2 MCS11 match for any gNB-UE distance when using HARQ-CC (see Table~\ref{tab_fixmcs1}), because they correspond to the same modulation order and ECR. However, under HARQ-IR setting, 
for distance of 50m, small differences appear for Table1 MCS13 and Table2 MCS7 because the effective ECR varies with retransmissions (see Eq.~\eqref{ecr} and its implementation details in Section~\ref{sec:tables}). Note that the effective ECR for MCS7 in Table2 can be reduced up to 0.37, while in the case of MCS13 in Table1, since it can be reduced up to 0.33, this provides slightly better results. 

Finally, note that a fixed MCS strategy (with the MCSs we have evaluated) cannot manage to serve a UE at 70m distance because of using a transmission rate much beyond the channel capacity, where even with HARQ combining, PHY data failures cannot be recovered.

From the results in Table~\ref{tab_fixmcs2}, which correspond to different MCSs of the same MCS Table2, 
it can be observed that, for 30m distance, a lower delay is attained at a lower MCS index due to the fact that retransmissions start earlier with a higher MCS index (see higher percentage of PHY data failures).
This is observed both for HARQ-CC and HARQ-IR. Indeed, the same effect is further aggravated for 50m distance, for which MCS11 is not adequate because of using a transmission rate beyond the channel capacity. For MCS11 and 30m distance, the effectiveness of HARQ-IR over HARQ-CC is observed again (as shown for MCS7 and 50m distance).

\subsection{Results with adaptive MCS}
\label{sec:res2}
In the second simulation campaign, we use link adaptation. 
Two link adaptation approaches are compared: 
\begin{itemize}
    \item Error model-based approach, in which the MCS is selected to meet a target transport BLER of at most 0.1 (see description in Section~\ref{sec:mcs})~\cite{mezzavilla:12}, 
    \item Shannon bound-based approach, which chooses the highest MCS that gives a spectral efficiency lower than the one provided by the Shannon rate (using a coefficient of ${-}\ln(5{\times}0.00001)/0.5$ to account for the difference in between the theoretical bound and real performance)~\cite{piro:11}.
\end{itemize}

The configuration is the same as for the previous simulation campaign, but, for this case, channel updates are disabled to avoid the different fading scenario for every packet and thus to enable proper link adaptation.
In Table~\ref{tab_amc1} and Table~\ref{tab_amc2}, we show the results for different NR MCS tables and HARQ combining methods, when using link adaptation based on the error model (Table~\ref{tab_amc1}) and based on the Shannon bound (Table~\ref{tab_amc2}). In this campaign, we also show the MCS that is selected for most of the transmissions in each configuration.

From the tables, it is observed that none of the methods lose any packet at application layer up to a distance of 70m, 
due to the combination of link adaptation and HARQ combining. This improves the end-to-end performance as compared to a fixed MCS strategy (see Tables~\ref{tab_fixmcs1}-\ref{tab_fixmcs2}). At 70m distance, due to the poor channel quality, failures at PHY layer occur, but they are completely recovered by HARQ. 

Results show that MCS selection is independent of the HARQ method. For selecting a MCS, we observe that the Shannon bound-based approach is more conservative than the Error model-based approach; the selected MCSs are lower than for the Error model-based approach. 
For that reason, in the Shannon bound-based approach, 0$\%$ packet losses are observed at PHY in all cases. 
The aggressiveness of the Error model-based approach, in turn, is shown to be beneficial for 30m distance, for which the delay is slightly reduced because less symbols are needed to complete a transmission. 
Note that in Error model-based approach (Table~\ref{tab_amc1}), at PHY layer, instead of 10$\%$ (i.e., the BLER constraint), 0$\%$ to 4$\%$ packet failures are obtained. This is because the SINR-BLER curves are quantized, the simulated CBS values in LLS are discrete, and due to the sub-optimality in MCS selection introduced by the 4-bit CQIs and the 5-bit MCSs defined in NR. Based on that, the selected MCS for each transmission provides a BLER lower than 0.1.

\section{Conclusions}
\label{sec:conc}
In this paper, we have presented an NR PHY abstraction model for system-level simulations of 5G networks. The proposed model is based on the EESM method and considers the latest NR specification regarding channel coding, LDPC base graph selection, code block segmentation, MCSs (for NR MCS Table1 and Table2), and HARQ combining methods. We have calibrated the EESM method with an NR-compliant LLS, from which we also obtained the SINR-BLER lookup tables for various settings of MCSs and resource allocations. Thanks to the calibration procedure, we found that the L2SM is insensitive to the NR numerology. Such a conclusion generalizes the application of the model to all NR numerologies and frequency ranges.
Finally, we have integrated the proposed PHY abstraction model into an ns-3 based NR SLS and illustrated its usage through system-level simulations by using different MCS tables, HARQ methods, and link adaptation approaches. The model is openly available to the research community, as a new interface of the ns-3 NR SLS, thus offering opportunities for reproducible research and collaborative model development.

\section{Acknowledgments}
\small
This work was partially funded by Spanish MINECO grant TEC2017-88373-R (5G-REFINE), Generalitat de Catalunya grant 2017 SGR 1195, and InterDigital Communications, Inc.

\bibliography{references}
\bibliographystyle{ieeetr}

\end{document}